\documentclass[aps,prd,superscriptaddress,twocolumn,nofootinbib]{revtex4-1}
\usepackage[utf8]{inputenc}
\usepackage{dsfont}
\usepackage{amsfonts}
\usepackage{amsmath}
\allowdisplaybreaks[4]        
\usepackage{amssymb}
\usepackage{euscript}     
\usepackage{braket}
\usepackage{starfont}
\usepackage{color,soul}         
\usepackage{tensor}        
\usepackage{amsthm}
\usepackage{graphicx}
\usepackage{slashed}
\usepackage{leftidx}
\usepackage{subfigure}
\usepackage{bbm}
\definecolor{outerspace}{rgb}{0.25, 0.29, 0.3}
\definecolor{scarlet}{rgb}{1.0, 0.13, 0.0}
\usepackage[header,title,page,titletoc]{appendix}  
\definecolor{princetonorange}{rgb}{1.0, 0.56, 0.0}
\definecolor{WildStrawberry}{rgb}{1.0, 0.26, 0.64}
\definecolor{rossocorsa}{rgb}{0.83, 0.0, 0.0}
\definecolor{navyblue}{rgb}{0.0, 0.0, 0.5}
\usepackage{float}
\usepackage[paper=letterpaper,margin=1in]{geometry}
\usepackage{mathtools}

\DeclareMathAlphabet{\pazocal}{OMS}{zplm}{m}{n}
\newcommand{\req}[1]{(\ref{#1})} 

\newcommand{\bea}{\begin{eqnarray}}
\newcommand{\diff}{\mathrm{d}}
\newcommand{\eea}{\end{eqnarray}}
\newcommand{\ba}{\begin{eqnarray}}
\newcommand{\ea}{\end{eqnarray}}

\newcommand{\be}{\begin{equation}}
\newcommand{\ee}{\end{equation} }
\newcommand{\beqa}{\begin{eqnarray}}
\newcommand{\eeqa}{\end{eqnarray}}
\newcommand{\beqar}{\begin{eqnarray*}}
\newcommand{\eeqar}{\end{eqnarray*}}

\renewcommand{\req}[1]{(\ref{#1})}

\newcommand{\eg}{{\it e.g.,}\ }

\makeatletter
\newcommand{\dal}{\mathop{\mathpalette\dal@\relax}}
\newcommand{\dal@}[2]{%
  \begingroup
  \sbox\z@{$\m@th#1\square$}%
  \dimen0=\fontdimen8
    \ifx#1\displaystyle\textfont\else
    \ifx#1\textstyle\textfont\else
    \ifx#1\scriptstyle\scriptfont\else
    \scriptscriptfont\fi\fi\fi3
  \makebox[\wd\z@]{%
    \hbox to \ht\z@{%
      \vrule width \dimen0
      \kern-\dimen0
      \vbox to \ht\z@{
        \hrule height \dimen0 width \ht\z@
        \vss
        \hrule height 2\dimen0
      }%
      \kern-2.5\dimen0
      \vrule width 2.5\dimen0
    }%
  }%
  \endgroup
}
\makeatother

\usepackage{hyperref}
\hypersetup{
    colorlinks,
    citecolor=blue,
    filecolor=navyblue,
    linkcolor=navyblue,
    urlcolor=navyblue
}

\begin{document}

\title{On the stability of Einsteinian Cubic Gravity black holes in EFT}
\author{Pablo Bueno}
\email{pablobueno@ub.edu}
\affiliation{Departament de F\'isica Qu\`antica i Astrof\'isica, Institut de Ci\`encies del Cosmos\\
 Universitat de Barcelona, Mart\'i i Franqu\`es 1, E-08028 Barcelona, Spain }

\author{Pablo A. Cano}
\email{pabloantonio.cano@kuleuven.be}
\affiliation{Instituut voor Theoretische Fysica, KU Leuven\\ Celestijnenlaan 200D, B-3001 Leuven, Belgium}

\author{Robie A. Hennigar}
\email{robie.hennigar@icc.ub.edu}
\affiliation{Departament de F\'isica Qu\`antica i Astrof\'isica, Institut de Ci\`encies del Cosmos\\
 Universitat de Barcelona, Mart\'i i Franqu\`es 1, E-08028 Barcelona, Spain }


\begin{abstract}
In this note we revisit the analysis performed in \cite{DeFelice:2023vmj} of odd-parity perturbations around static and spherically symmetric black holes in Einsteinian Cubic Gravity (ECG). We show that the additional propagating modes always have masses much above the cutoff of the theory. Therefore, contrary to what is claimed in that paper,  the ECG black holes remain stable within the effective field theory regime. We consider the same analysis for a general cubic theory, showing that the ECG results are not special in this regard. We use the occasion to make some clarifications on the role, uses and limitations of ECG and its generalizations.
\end{abstract}
\maketitle

\section{Einsteinian Cubic Gravity:\\ what it is and what it is not}
Einsteinian Cubic Gravity (ECG) was originally constructed in \cite{Bueno:2016xff} as the most general gravity action including terms with up to six derivatives of the metric built from contractions of the Riemann tensor and the metric such that: i) its linearized equations on general maximally symmetric backgrounds are equivalent, up to a renormalization of the Newton constant, to the Einstein gravity ones; ii) at each order, the relative coefficients of the curvature invariants involved are the same in general dimensions. 

These criteria select a linear combination of 
 the cosmological constant, the Einstein-Hilbert action, the quadratic and cubic Lovelock densities, plus a new cubic density given by\footnote{There are many developments related to ECG and its generalizations which have not been included in the present summary ---these are contained in \cite{Bueno:2016xff,Dey:2016pei,Hansen:2016gud,Hennigar:2016gkm,Bueno:2016lrh,Bueno:2016ypa,Chernicoff:2016qrc,Cisterna:2017umf,Ghodsi:2017iee,Dykaar:2017mba,Hennigar:2017ego,Bueno:2017sui,Ahmed:2017jod,Bueno:2017qce,Ortin:2017fux,Hennigar:2017umz,Feng:2017jub,Feng:2017tev,Li:2017ncu,Concha:2017nca,Bueno:2017ctd,Adami:2017phg,Li:2017txk,Hennigar:2018hza,Myung:2018ete,Bueno:2018xqc,Peng:2018vbe,Myung:2018qsn,Chernicoff:2018apt,Jiang:2018sqj,Li:2018drw,Bueno:2018uoy,Bueno:2018yzo,Hennigar:2018cnh,Cisterna:2018jqg,Arciniega:2018fxj,Poshteh:2018wqy,Cisterna:2018tgx,Carballo-Rubio:2018bmu,Arciniega:2018tnn,GuajardoR:2019qkx,Cano:2019ore,Li:2019auk,Lu:2019urr,Mir:2019ecg,Mir:2019rik,Jiang:2019fpz,Mehdizadeh:2019qvc,Erices:2019mkd,Bahamonde:2019shr,Emond:2019crr,Li:2019tpf,Bueno:2019ltp,Cristofoli:2019ewu,Arciniega:2019oxa,Moynihan:2019bor,Bueno:2019ycr,CanoMolina-Ninirola:2019uzm,Jiang:2019kks,Cano:2019ozf,Burger:2019wkq,Burger:2020tri,Giribet:2020aks,Bueno:2020odt,Frassino:2020zuv,KordZangeneh:2020qeg,Marciu:2020ysf,Quiros:2020uhr,Pookkillath:2020iqq,Marciu:2020ski,Adair:2020vso,Liu:2020yqa,Khodabakhshi:2020hny,Edelstein:2020nhg,Konoplya:2020jgt,Khan:2020kwl,Khodabakhshi:2020ddv,Cano:2020qhy,Cano:2020ezi,Quiros:2020eim,Edelstein:2020lgv,Do:2020vdc,BeltranJimenez:2020lee,RuiperezVicente:2020qfw,Caceres:2020jrf,Colleaux:2020wfv,Mustafa:2020qjo,Cano:2020oaa,Fierro:2020wps,Bueno:2020uxs,Pokrovsky:2021flw,VilarLopez:2021ebu,Alarcon:2021tda,JimenezCano:2021rlu,Anastasiou:2021swo,Bhattacharjee:2021nfx,Bhattacharjee:2021jwm,Jawad:2021kkp,Bueno:2021krl,Anastasiou:2021tlv,Giri:2021amc,Cristofoli:2021vqw,Araya:2021atx,Li:2021jfh,Knorr:2021lll,Ghosh:2021zpb,Bakhtiarizadeh:2021vdo,Sardar:2021blt,Asimakis:2021yct,Bakhtiarizadeh:2021hjr,Saha:2022vcb,Jaime:2022cho,Bueno:2022lhf,Edelstein:2022xlb,Cano:2022ord,Kluth:2022vnq,Hussain:2022lxb,Marciu:2022wzh,Bueno:2022jbl,Bueno:2022res,Erices:2022bws,Edelstein:2022lco,Sultan:2022opy,Rudra:2022qbv,Marciu:2022rsc,Sajadi:2022pcz,Agurto-Sepulveda:2022vvf,Moreno:2022unk,Sajadi:2022ybs,Delhom:2022vae,Bazeia:2022agk,Gil:2022ubv,PereniguezRodriguez:2022eal,Sultan:2022aoa,Bueno:2022ewf,Jawad:2022hgk,Mazloumi:2022nvi,Asimakis:2022mbe,Chen:2022fdi,Rani:2023wco,Hu:2023juh,Marks:2023ipa,Murcia:2023zok,Lobao:2023wde,Cano:2023dyg,Mustafa:2023vvt,Do:2023yvg,Moreno:2023rfl,Singha:2023bon,DeFelice:2023vmj,Peng:2023zvf,Lessa:2023xto}.}
\begin{align}
\mathcal{P}\equiv  &+12 \tensor{R}{_{a}^{c}_{b}^{d}}\tensor{R}{_{c}^{e}_{d}^{f}}\tensor{R}{_{e}^{a}_{f}^{b}}+R_{ab}^{cd}R_{cd}^{ef}R_{ef}^{ab}\\ \notag &-12R_{abcd}R^{ac}R^{bd}+8R_{a}^{b}R_{b}^{c}R_{c}^{a}\, .
\end{align}
In four spacetime dimensions, the  quadratic and cubic Lovelock pieces become topological and trivial respectively, and one is left with a linear combination of the Einstein-Hilbert term and $\mathcal{P}$ of the form
\begin{equation}\label{action}
S=\frac{1}{16\pi G}\int \diff^4x\sqrt{|g|}\left[R-\frac{\mu \ell^4}{8}\mathcal{P}\right]\, ,
\end{equation}
where $\mu$ is a dimensionless coupling and we have set the cosmological constant to zero (although this is not necessary).

Soon after the theory was constructed, it was observed that it possesses an interesting additional feature in the four-dimensional case. Namely, it admits  static and spherically symmetric black hole solutions which are continuous generalizations of the Schwarzschild one, characterized by a single metric function \cite{Hennigar:2016gkm,Bueno:2016lrh}. The solutions have the form
\begin{equation}\label{sf}
\diff s ^2=- f(r) \diff t ^2 +\frac{\diff r^ 2}{f(r)}+r^ 2 (\diff \theta^2+\sin^2\theta \diff \phi^2)\, ,
\end{equation}
where the metric function satisfies a second-order differential equation given by
\begin{align}
\begin{split}
&1-\frac{2GM}{r}=f+\frac{\mu \ell^4}{8r }\left[ 
4f'^3+\frac{12f'^2}{r} \right. \\ &\left. -24f(f-1)\frac{f'}{r^2}-12f f''\left(f'-\frac{2(f-1)}{r}\right)\right]\, ,
\end{split}
\end{align}
where $M$ is an integration constant which will correspond to the mass of the solutions. The above equation cannot be solved analytically, but it is still easy to verify that there exists a unique mass-$M$ black hole solution with the right asymptotic behavior whenever $\mu \geq 0$. At leading order in $\mu$, the solution takes the form
\begin{equation}\label{mf}
f=1-\frac{2GM}{r}- \left[ \frac{54(GM)^2}{r^6}-\frac{92 (GM)^3}{r^7}  \right]  \mu \ell^ 4\, .
\end{equation}
Near-horizon and asymptotic expansions can be easily obtained \cite{Bueno:2016lrh}, as well as an analytic approximation in terms of continued fractions \cite{Hennigar:2018hza}. The full solution for fixed values of $ \mu \ell^ 4$ can be constructed numerically.

These results are surprising, as generic higher-derivative gravities require $g_{tt} g_{rr}\neq -1$. On the other hand, examples of theories satisfying this property were known to exist in $D\geq 5$: Lovelock theories on the one hand \cite{Lovelock1,Wheeler:1985nh,Boulware:1985wk,Cai:2001dz}, and the so-called Quasi-topological gravities \cite{Oliva:2010eb,Quasi,Dehghani:2011vu,Cisterna:2017umf}, on the other. The observation that a four-dimensional theory admitting solutions of this kind existed, triggered the discovery of a new class of theories \cite{Hennigar:2017ego,Bueno:2017sui} which in fact contained all the aforementioned cases. These ``Generalized Quasi-topological gravities'' (GQTs) are defined by the following technical condition: evaluating the gravitational Lagrangian on an ansatz of the form (\ref{sf}), we say the theory is of the GQT class if the Euler-Lagrangian equation resulting from varying it with respect to $f(r)$ vanishes identically ---see \eg \cite{Bueno:2022res} for more details. When this condition holds, the theory admits solutions of the form (\ref{sf}) where  the equation of $f(r)$ can always be integrated once, resulting in a equation which is either algebraic or differential of degree 2.\footnote{If covariant derivatives of the Riemann tensor appear in the action, this degree can in principle be higher, but this has not been explored so far.} The former case corresponds to Quasi-topological and Lovelock theories, and only occurs in $D\geq 5$, whereas the latter occurs in $D\geq 4$, with ECG being the lowest-order example in four dimensions. It is possible to prove that GQTs have linearized equations identical to the Einstein gravity ones on maximally symmetric backgrounds \cite{Bueno:2017sui}, but observe that the converse implication is not true (namely, there exist theories with Einstein-like linearized equations which are not GQTs).

The GQTs static and spherically symmetric black holes are \emph{bona-fide} continuous generalizations of the Schwarzschild one controlled by the corresponding gravitational couplings. In particular, as shown in \cite{Bueno:2017sui}, they are the solutions representing the gravitational field outside spherically symmetric distributions of mass $M$ in those theories. This is nontrivial and, in fact, it is possible to argue that only theories possessing the same linearized spectrum as Einstein gravity are susceptible of  admitting solutions characterized by a single function which describe the exterior field of spherically symmetric mass distributions. In particular, this implies that the Schwarzschild metric cannot describe the exterior field of a spherical body in generic $f(R)$ gravities, even though it is a solution of those theories.

Additional four-dimensional black hole solutions with $g_{tt}g_{rr}\neq 1$ are known to exist for Einstein gravity plus quadratic terms \cite{Lu:2015psa,Lu:2015cqa}. However, 
such black holes do not have an interpretation in the  effective field theory
 (EFT) context, since one can remove all quadratic terms from the action using field redefinitions.\footnote{This is because the most general non-trivial (or topological) quadratic action in $D=4$ involves a linear combination of $R^ 2$ and $R_{ab}R^{ab}$ terms and, on general grounds, any Ricci tensors can be removed order by order in the EFT by redefinitions of the form $g_{ab} \rightarrow g_{ab}+ \alpha R_{ab}$ for certain constant $\alpha \ll 1$ ---see \eg \cite{Bueno:2019ltp}.}

While the order-reduction phenomenon of the equations of motion occurring for static GQT black holes has been observed to extend in some cases to more general backgrounds ---including: extremal rotating black holes \cite{Cano:2019ozf}, slowly rotating black holes \cite{Adair:2020vso}, Taub geometries \cite{Bueno:2018uoy} and cosmological spacetimes \cite{Arciniega:2018fxj,Arciniega:2018tnn,Cisterna:2018tgx}--- of all GQTs, only  the Lovelock ones have second-order equations on general backgrounds. Consequently, 
one expects stability issues to arise in certain regimes for these theories, just like for any generic higher-curvature gravity. For instance, any String Theory effective action truncated at a finite order in $\alpha'$ will have the same issue.

The structure and types of GQTs in general dimensions have been studied in detail in \cite{Bueno:2019ycr,Bueno:2022res,Moreno:2023rfl}. In particular, it has been shown that GQT densities exist for all $D\geq 4$ and for arbitrarily high curvature orders. Importantly, it was proven in \cite{Bueno:2019ltp} that \emph{any} higher-curvature gravitational effective action involving densities built from the Riemann tensor and the metric can be mapped, by a field redefinition, to a GQT. In other words, GQTs provide a generating set of invariants for the most general gravitational action. In particular, ECG captures the most general effective field theory (EFT) extension of Einstein gravity up to six derivatives. One of the advantages of working in the GQTs frame is that black hole solutions are much simpler and universal, while the physical properties remain the same ---see \cite{Bueno:2019ltp} for an explicit example. Observe that this implies, in particular, that if there was an inconsistency with the solutions of ECG or GQTs in the perturbative regime beyond Einstein gravity, this would spoil the validity of perturbative modifications of the Schwarzschild black hole for completely general higher-curvature actions. This seems unlikely. As we show below, the arguments presented in \cite{DeFelice:2023vmj} suggesting this to be the case are wrong.



GQT black holes have also been widely studied beyond the EFT regime. Let us say a few words about this. Firstly, a particularly nice feature of these solutions is that many of their physical properties can be evaluated analytically for non-perturbative values of the gravitational couplings ---even when the solutions themselves cannot. In particular, this gives rise to a procedure which allows to test the effects of infinite towers of higher-derivative terms ---see \eg \cite{Bueno:2017qce} for black hole thermodynamics and \cite{Arciniega:2018tnn} for early-time cosmological evolution--- by considering the effect of each of those terms at arbitrary order and resuming them all. The resumed result can be hoped to capture some of the effects present in a full quantum gravity calculation. This idea certainly departs from the standard EFT approach and the validity of its conclusions is unclear. Still, it is one of the only setups ---see \cite{Hohm:2019jgu} for another--- in which questions like this can be formulated and answered explicitly.


From a different perspective, higher-curvature gravities provide toy models of holographic CFTs inequivalent to Einstein gravity ---see \eg \cite{Buchel:2009sk,Myers:2010jv}. This has been exploited to study many physical properties, and to discover  various universal relations valid for completely general theories \cite{Brigante:2007nu,Bueno:2022jbl,Myers:2010tj,Perlmutter:2013gua,Mezei:2014zla,Bueno:2015rda,Chu:2016tps}. GQTs are particularly useful for this, as many of the holographic quantities which can be accessed using black holes and other solutions are much easier to obtain and can be extracted fully analytically. In particular, $D=4$ ECG provides a somewhat canonical toy model of a three-dimensional CFT with a non-vanishing stress-tensor three-point function coefficient $t_4$ \cite{Bueno:2018xqc,Edelstein:2022xlb}. This has been used to obtain a universal formula for the partition function of general three-dimensional CFTs on a slightly squashed sphere \cite{Bueno:2018yzo,Bueno:2020odt}. It is conceivable that additional universal results can be obtained from holographic GQTs in the future.



\section{Odd parity perturbations of ECG static black holes}

As we have mentioned in the introduction, ECG captures the most general EFT extension of vacuum general relativity up to six derivatives. In this context, the dimensionful parameter $\ell$ appearing in the action~\eqref{action} naturally represents the cutoff of the theory, in the sense that one should consider curvatures much smaller than $1/\ell^2$ in order to ensure that the higher-derivative terms remain small. In the  same way, propagating modes with momentum $k$ should also satisfy $|k| \ll 1/\ell$. Any mode of the EFT with momentum above this cutoff is unphysical as at this point the massive modes that have been integrated out from the putative UV complete theory enter into play.  

Keeping these ideas in mind, let us now turn to an analysis of the perturbations of the static and spherically symmetric black holes with metric and metric function given by~\eqref{sf} and~\eqref{mf}. Note that this solution is within the EFT regime provided that the size of the black hole is much larger than the length scale $\ell$,
\begin{equation}
M\sim r_{h}\gg \ell\, ,
\end{equation}
which we shall assume throughout the rest of the paper.

In order to fix a few concepts, it is useful to study first the perturbations of a massive test scalar field on the background of this black hole. Thus, we start with the Klein-Gordon equation
\begin{equation}
\nabla^2\varphi=\mathcal{M}^2 \varphi\, ,
\end{equation}
and we separate the scalar field as
\begin{equation}
\varphi=e^{-i\omega t} \Psi(r) Y_{lm}(\theta,\phi)\, ,
\end{equation}
obtaining the equation
\begin{equation}
\frac{\omega^2}{f(r)}+\frac{1}{r^2}\frac{\diff }{\diff r}\left(r^2f(r)\Psi'(r)\right)-\frac{l(l+1)}{r^2}\Psi(r)=\mathcal{M}^2\Psi(r)\, .
\end{equation}
We are interested in modes of large momentum and therefore let us write
\begin{equation}
\Psi(r)=A(r)e^{i k_{r} r_*}\, ,
\end{equation}
where $r_*$ is the tortoise coordinate defined by $\diff r_*=\diff r/f(r)$, and the amplitude $A(r)$ is supposed to remain approximately constant in distances of the order of $1/k_{r}$, this is, 
\begin{equation}
\left|\frac{A'}{A}\right| \ll |k_{r}|\, .
\end{equation}
Then, in the limit of large $k_{r}$ and large $l$ we get the following dispersion relation for this mode, 
\begin{equation}
\omega^2=k_{r}^2+f k_{\theta}^2+f \mathcal{M}^2\, ,
\end{equation}
where we have introduced the notation
\begin{equation}
k_{\theta}=\frac{l}{r}\, .
\end{equation}
For $\mathcal{M}=0$, this mode moves at the speed of light, but more generally we could have dispersion relations of the form
\begin{equation}\label{generaldisp}
\omega^2=c_{r}^2k_{r}^2+c_{\theta}^2f k_{\theta}^2+f \mathcal{M}^2\, ,
\end{equation}
where $c_{r}$ and $c_{\theta}$ represent the speed of propagation in the radial and angular directions in the limit of infinite momentum. The dispersion relations we obtain for gravitational perturbations of ECG take precisely this form.

Let us then consider odd-parity gravitational perturbations on top of the SSS metric. In the Regge-Wheeler gauge, the perturbations with harmonic numbers $(l,m)=(l,0)$ can be written as 
\begin{equation}
h_{\mu\nu}=e^{-i\omega t}\begin{pmatrix}
0&0&0&h_0(r)\\
0&0&0&h_1(r)\\
0&0&0&0\\
h_0(r)&h_1(r)&0&0
\end{pmatrix}\sin\theta \frac{\partial Y_{l0}}{\partial \theta}\, .
\end{equation}
Note that there is no loss of generality in setting $m=0$ as the equations for the radial functions $h_{0,1}$ are independent of $m$.  We perform a direct evaluation of the equations of motion at linear order in $h_{\mu\nu}$ (instead of using the reduced Lagrangian). The relevant components of the equations of motion are $\mathcal{E}_{r\phi}$ (of fourth order) and $\mathcal{E}_{\theta\phi}$ (of third order). These take a complicated form, so, as in \cite{DeFelice:2023vmj}, we consider solutions of large momentum. A convenient way of writing the radial functions is as follows
\begin{equation}
h_{0}(r)=A_0(r) e^{i k_{r} r_*}\, ,\quad h_{1}(r)=\frac{A_1(r)}{f(r)}e^{i k_{r} r_*}\, ,
\end{equation}
where, as before $r_*$ is the tortoise coordinate, and the amplitudes $A_{0,1}(r)$ are supposed to vary slowly,
\begin{equation}
\left|\frac{A_{i}'}{A_{i}}\right|\ll |k_{r}|\, .
\end{equation}
The factor of $1/f(r)$ in $h_1$ is introduced to get a well-behaved perturbation at the horizon. 

We then consider the large momentum/frequency limit. It must be emphasized that this limit has to be taken with care. If the large momentum limit is taken naively, then the higher-derivatives terms dominate when we take $\omega\rightarrow\infty$, $k\rightarrow\infty$, and thus this is outside of the EFT regime. Therefore, we should choose a momentum that is much larger than the inverse of the size of the black hole but much smaller than the cutoff of the theory. This is, 
\begin{equation}
\frac{1}{r_{h}}\ll |k| \ll \frac{1}{\ell}\, .
\end{equation} 
In practice, this means that we must take the limit independently in the Einstein and in the higher-derivative contributions to the equations of motion, keeping only the leading terms in each case. Proceeding in this way, we get 
\begin{widetext}
\begin{align*}
\mathcal{E}_{\theta\phi}\propto &\left(A_0 \omega +A_1 k_r\right) \left(4 f r+3 \mu \ell^4 \left(-2 \omega ^2+2 k_r^2-f k_{\theta
   }^2\right) \left(-f'+r f''\right)\right)\, ,\\
   \mathcal{E}_{r\phi}\propto&-8 r^5 \left(\omega  \left(\omega  A_1+A_0 k_r\right)-f A_1 k_{\theta }^2\right)+\frac{6 r^3 \mu \ell^4}{f} 
   \left[\omega  \left(\omega  A_1+A_0 k_r\right) \left(\omega ^2-k_r^2\right) \left(2+r f'\right)\right.\\
   &\left.+f^2
   k_{\theta }^2 \left(2 \omega  \left(\omega  A_1+A_0 k_r\right)+r A_1 k_{\theta }^2 f'-r^2 A_1 k_{\theta
   }^2 f''\right)\right.\\
   &\left.+f \left(-2 \omega  \left(\omega  A_1+A_0 k_r\right) \left(\omega ^2-k_r^2+k_{\theta
   }^2\right)-2 r \left(2 \omega  A_0 k_r+A_1 \left(\omega ^2+k_r^2\right)\right) k_{\theta }^2 f'\right.\right.\\
   &\left.\left.+r^2
   \left(\omega ^2 A_1+3 \omega  A_0 k_r+2 A_1 k_r^2\right) k_{\theta }^2 f''\right)\right]\, .
   \end{align*}
   \end{widetext}
Let us analyze the solutions to these equations. There are three possible dispersion relations that can be obtained, and we discuss each in turn.

\subsection*{Solution 1}
The equation $\mathcal{E}_{\theta\phi} = 0$ can be solved by 
\begin{equation}
A_{0}=-A_{1}\frac{k_r}{\omega}\, .
\end{equation}
Plugging this into the second equation, we obtain
\begin{equation}
\begin{aligned}\label{subbed}
&\left(-\omega ^2+k_r^2+f k_{\theta }^2\right) \left[4 f r^2+\mu  \left(3 \left(-\omega ^2+k_r^2\right)
   \left(2\right.\right.\right.\\
   &\left.\left.\left.+r f'\right)+f \left(6 \omega ^2-6 k_r^2+3 r k_{\theta }^2 f'-3 r^2 k_{\theta }^2
   f''\right)\right)\right]=0\, .
   \end{aligned}
\end{equation}
Solving the first factor gives the dispersion relation
\begin{equation}
\omega^2=k_{r}^2+f k_{\theta}^2\, ,
\end{equation}
which is the standard result in Einstein gravity. 

\subsection*{Solution 2}

Just like for Solution 1, we take 
\begin{equation}
A_{0}=-A_{1}\frac{k_r}{\omega}\, ,
\end{equation}
which solves $\mathcal{E}_{\theta\phi} = 0$. Now we solve the second factor in~\eqref{subbed} which gives the following dispersion relation 
\begin{equation}
\omega^2=k_r^2+k_{\theta }^2\frac{f r \left(f'-r f''\right)}{2-2 f+r f'}+\frac{4 r^2 f}{3 \mu \ell^4  \left(2-2f+r f'\right)}\, .
\end{equation}
Thus, comparing with \req{generaldisp} we find
\begin{equation}
\begin{aligned}
c_{r}^2=&1\, ,\\
c_{\theta}^2=&\frac{r \left(f'-r f''\right)}{2-2 f+r f'}\approx 1 \, ,\\
\mathcal{M}^2=&\frac{4 r^2}{3 \mu \ell^4  \left(2-2f+r f'\right)}\approx \frac{2 r^3}{9\mu \ell^4 M} \, .
\end{aligned}
\end{equation}
Note that for $r>r_{h}$ we necessarily have $\mathcal{M}^2\gg 1/\ell^2$, and hence according to our previous discussion this mode is unphysical and should be regarded as an artifact of the EFT.

\subsection*{Solution 3}
Returning to  $\mathcal{E}_{\theta\phi}$, we can instead solve the second factor of this equation to obtain the third dispersion relation,
\begin{equation}
\omega^2=k_r^2-\frac{f}{2}k_{\theta}^2-\frac{2 r f}{3 \mu \ell^4( f'- r f'')}\, ,
\end{equation}
and now we identify
\begin{equation}
\begin{aligned}
c_{r}^2=&1\, ,\\
c_{\theta}^2=&-\frac{1}{2} \, ,\\
\mathcal{M}^2=&-\frac{2 r }{3 \mu \ell^4( f'- r f'')}\approx-\frac{r^3}{9 \mu \ell^4M} \, .
\end{aligned}
\end{equation}
Again this is a massive mode with a mass above the cutoff and therefore it is not meaningful within EFT. Observe that if one naively takes $k_{\theta}\rightarrow \infty$, as in \cite{DeFelice:2023vmj}, one would conclude that this mode is unstable as it has a negative speed squared in the angular direction. However, this requires taking $k_\theta$  above the cutoff $1/\ell$, and hence the conclusion is not valid.

\subsection{Extra modes and validity of the EFT}

Of the above modes, only that described by Solution 1 is within the domain of validity of the EFT. Solutions 2 and 3 both have masses much above the cutoff, as we now describe in more detail. The most natural choice for the cutoff is the scale given by the coupling constant of the theory $\ell_{\rm cutoff} \sim \mu^{1/4} \ell$. Without loss of generality, we can take $\mu$ to be an $\mathcal{O}(1)$ constant so that information about the cutoff is contained in $\ell$. Then, the requirement that these modes be within the EFT approximation requires that
\be 
\mathcal{M}^2 \ll \frac{1}{\ell^2} \, \, \Rightarrow \, \, r^3 \ll \mu \ell^2 r_h \, .
\ee 
Now combining this with the requirement that the corrections to the metric should remain small, i.e. $r_h \gg \ell$, we have the chain of inequalities
\be 
r^3 \ll \mu \ell^2 r_h  \ll r_h^3 \, .
\ee
Thus, we see that this condition can only be satisfied for $r \ll r_h$ deep within the black hole. These modes are therefore not relevant for an EFT description of the physics outside the black hole horizon. Note also that this conclusion is independent of these modes being tachyonic or not ---in fact, one of the extra modes must always be a tachyon, $\mathcal{M}^2<0$. This is because, having $|\mathcal{M}^2|\gg 1/\ell^2$, such mode would have a frequency/momentum larger than the cutoff in the black hole exterior, and therefore it lies beyond the limit of validity of EFT.

Actually, the authors of \cite{DeFelice:2023vmj} do realize that their conclusions are only valid when their equation (3.32) is satisfied,
\be 
\frac{2r_h^2}{9 \mu \ell^4 } \frac{r^3}{r_h^3} \ll M^2_{\rm cutoff} \, .
\ee
In our notation, this is equivalent to the condition $\mathcal{M}^2\ll 1/\ell^2$. They do not realize however that if the cutoff of the theory coincides with the scale of the coupling constant of the higher derivative terms (as we have just described) then their equation (3.32) is \emph{never} satisfied. An alternative way to phrase this conclusion is that rather than excluding the ECG black holes, the authors of~\cite{DeFelice:2023vmj} instead exclude cutoffs for the theory at energy scales that are parametrically smaller than the natural cutoff set by the coupling. 

In fact, none of these results are particular to ECG. In the appendix of this manuscript we have considered odd-parity gravitational perturbations for a theory consisting of the most general extension of general relativity with cubic powers of the curvature. Qualitatively the analysis is exactly the same. There is a single mode that is within the domain of validity of the EFT, while two additional modes have masses that are much above the cutoff of the theory. This is not surprising. As we have emphasized in the introduction, ECG captures the most general six-derivative corrections to Einstein gravity within the EFT regime. Had our analysis concluded something special for ECG in the regime where EFT is valid, it would have signalled issues for gravitational EFT more generally. 

\section{Discussion}
Higher-derivative gravities, considered as EFT extensions of general relativity, generally lead to the propagation of additional degrees of freedom that can have pathological properties. However, all of these modes possess masses exceeding the theory's cutoff, rendering them outside the domain of validity of the EFT.

We have shown this to be the case for odd-parity perturbations around the spherically symmetric black hole solutions of ECG, but there is nothing special about this example, and we expect this conclusion to hold for general theories and solutions. 
Therefore, one should understand the extra modes and their associated instabilities as artifacts of the EFT rather than genuine physical degrees of freedom.  In particular, this implies that ECG black holes are physically meaningful as long as their radius is larger than the cutoff length scale $\ell$. 

From a broader perspective, we conclude that it should be possible to make sense of dynamical evolution of higher-curvature gravities within the EFT regime. How to excise the unphysical instabilities and extract the dynamics in a sensible way is of course a much more complicated question. However, there is compelling evidence that this is possible. Recent works have been able to perform numerical relativity simulations in some of these theories by reformulating the equations of motion in an appropriate way \cite{Cayuso:2020lca,Cayuso:2023aht}. 
This evidence strongly suggests that, despite their apparent pathological behavior, higher-curvature theories indeed give rise to well-behaved dynamics within the EFT regime.

In this context, ECG and the whole family of GQTs are no different. What makes these theories remarkable is that they allow us to answer many questions about black holes, holography, cosmology and other areas in a much simpler and clearer way.  Hence, ECG and its generalizations remain as useful theories that allow us to understand the effects of higher-curvature corrections across a wide range of scenarios. 
 


\vspace{0.4cm}
\begin{acknowledgments} 
PAC would like to thank Pau Figueras for enlightening discussions on higher-curvature theories and numerical relativity. 
RAH is grateful to the Department of Applied Mathematics and Theoretical Physics at the University of Cambridge for hospitality during the completion of this work and thanks Iain Davies and Harvey Reall for helpful discussions.
PB was supported by a Ram\'on y Cajal fellowship (RYC2020-028756-I) from Spain's Ministry of Science and Innovation. 
The work of PAC is supported by a postdoctoral fellowship from the Research Foundation - Flanders (FWO grant 12ZH121N). The work of RAH received the support of a fellowship from ``la Caixa” Foundation (ID 100010434) and from the European Union’s Horizon 2020 research and innovation programme under the Marie Skłodowska-Curie grant agreement No 847648 under fellowship code LCF/BQ/PI21/11830027.

\end{acknowledgments}

\appendix
\onecolumngrid \vspace{1.5cm}

\section{Odd Parity Perturbations for General Cubic Gravities}

In this appendix we extend the results of the main text to general theories that supplement Einstein gravity with terms cubic in the curvature. As we will see, the results are qualitatively no different than the ECG case treated in the main text. This highlights the fact that the issues claimed in~\cite{DeFelice:2023vmj} are not particular to ECG, and just like in that case, the purported pathologies are always beyond the EFT regime. 

We begin with the following action,
\begin{align} 
S=  \frac{1}{16\pi G}  \int  \diff^4x\sqrt{|g|}   \bigg[ &R + \ell^4 \bigg(c_1 \tensor{R}{_{a}^{c}_{b}^{d}}\tensor{R}{_{c}^{e}_{d}^{f}}\tensor{R}{_{e}^{a}_{f}^{b}} + c_2 \tensor{R}{_{ab}^{cd}} \tensor{R}{_{cd}^{ef}} \tensor{R}{_{ef}^{ab}} + c_3 \tensor{R}{_{abcd}} \tensor{R}{^{abc}_{e}} \tensor{R}{^{de}} 
\nonumber\\
&+ c_4 \tensor{R}{_{abcd}} \tensor{R}{^{abcd}} R+c_5 \tensor{R}{_{abcd}} \tensor{R}{^{ac}} \tensor{R}{^{bd}} + c_6 \tensor{R}{_{a}^{b}} \tensor{R}{_{b}^{c}} \tensor{R}{_{c}^{a}} + c_7 R_{ab} R^{ab} R + c_8 R^3  \bigg) \bigg]  \, ,
\end{align}
where as before $\ell$ is a length scale and $c_i$ are dimensionless coupling constants. Restricting to static and spherically symmetric metrics, the above action admits a solution of the form
\be 
\diff s^2 = - N(r)^2 f(r) \diff t^2 + \frac{\diff r^2}{f(r)} + r^2 \left( \diff  \theta^2 + \sin^2 \theta \diff \phi^2 \right) \, .
\ee
In this case, obtaining the solution involves solving a pair of coupled fourth-order non-linear ordinary differential equations for the metric functions $f(r)$ and $N(r)$. In general, this will be a formidable task and here we will content ourselves with a power-series solution constructed by solving the equations of motion perturbatively near $r \to \infty$. The result to $\mathcal{O}(\ell^4)$ is,
\begin{align} 
f(r) &= 1 - \frac{r_0}{r} + \frac{18 \ell^4 r_0^2 \left(3c_2 + c_3 + 4 c_4 \right)}{r^6} + \frac{\ell^4 r_0^3 \left(c_1 - 196 c_2 - 66 c_3 - 264 c_4 \right)}{4 r^7} + \cdots \, ,
\\
N(r) &= 1 + \frac{3 \ell^4 r_0^2 \left( 3 c_1 - 36 c_2 - 14 c_3 - 56 c_4 \right)}{4 r^6} + \cdots \, .
\end{align} 
Here we have normalized $N(r)$ such that $N(r) \to 1$ as $r \to \infty$. 

We wish to study odd-parity gravitational perturbations for these black holes. We proceed in the same manner as in the main text. We work in the Regge-Wheeler gauge, writing the perturbations as 
\begin{equation}
h_{ab}=e^{-i\omega t}\begin{pmatrix}
0&0&0&h_0(r)\\
0&0&0&h_1(r)\\
0&0&0&0\\
h_0(r)&h_1(r)&0&0
\end{pmatrix}\sin\theta \frac{\partial Y_{l0}}{\partial \theta}\, .
\end{equation}
As before, instead of Lagrangian methods, we perform a direct evaluation of the field equations for the perturbed metric. The relevant components of the field equations are $\mathcal{E}_{r\phi}$ and $\mathcal{E}_{\theta\phi}$ which are in general horribly complicated expressions. We seek solutions in the limit of large momentum, writing the radial functions as follows
\begin{equation}
h_{0}(r)=A_0(r) e^{i k_{r} r_*}\, ,\quad h_{1}(r)=\frac{A_1(r)}{f(r)}e^{i k_{r} r_*}\, , \quad r_* = \int \frac{\diff r}{N(r) f(r)} \, .
\end{equation} 
Taking the large momentum limit carefully we obtain the following equations that will determine the dispersion relation for the various modes,
\begin{align}
\mathcal{E}_{\theta \phi} \propto & +\left(\omega N A_0 + k_r A_1 \right) \bigg[2 r^2 f N^2 + \ell^4 N^2 f k_\theta^2 \left(\alpha r^2 f'' - 3 \beta r f' + 24 \nu (f-1) \right) 
\nonumber\\
& + 2 \ell^4 (N^2 \omega^2- k_r^2) \left( - (\alpha + \xi) r^2 f'' - (12 \nu + \alpha) r f' + (3 \beta + 2 \xi + 3 \alpha + 12 \nu)(f-1) \right) \bigg]\, ,
\\
\mathcal{E}_{r \phi} \propto &- 8 N^2 f r^5 \left(N(\omega^2 - f k_\theta^2) A_1 + k_r \omega A_0 \right) -4 \ell^4 r^3 \big(-A_0 \omega k_r \big(N^2 \big(f k_{\theta }^2 \big(r^2 f'' 
   (\alpha -12 \nu +2 \xi ) 
  \nonumber\\
  &+2 r f' (\alpha +3 (\beta +4 \nu ))-2 (f-1) (4 \alpha +3 \beta
   +24 \nu +2 \xi )\big)+\omega ^2 \big(12 \nu  r^2 f''-3 \beta  r f'+2 \alpha 
   (f-1)\big)\big)
   \nonumber\\
   &+k_r^2 \big(2 \big(\alpha -6 \nu  r^2 f''-\alpha  f\big)+3
   \beta  r f'\big)\big)-A_1 \big(N^3 \big(\omega ^2-f k_{\theta }^2\big)
   \big(f k_{\theta }^2 \big(-\alpha  r^2 f''+3 \beta  r f'-24 (f-1) \nu \big)
   \nonumber\\
   &+\omega^2 \big(12 \nu  r^2 f''-3 \beta  r f'+2 \alpha  (f-1)\big)\big)+N k_r^2
   \big(2 f k_{\theta }^2 \big(r^2 (\alpha +\xi ) f''+r (\alpha +12 \nu ) f'
   \nonumber\\
   &-(f-1) (3
   \alpha +3 \beta +12 \nu +2 \xi )\big)+\omega ^2 \big(2 \big(\alpha -6 \nu  r^2
   f''-\alpha  f\big)+3 \beta  r f'\big)\big)\big)\big) \, .
\end{align}
To simplify the expressions, we have introduced the following shorthand forms for certain combinations of the coupling constants:
\begin{align}
\alpha &\equiv c_3 + 8 c_4 + \frac{3}{2} c_6 + 2 c_7 \, ,
\\
\beta &\equiv  c_1 - 4c_2 - \frac{10}{3} c_3 - \frac{32}{3} c_4 - c_5 - 2 c_6 - \frac{8}{3} c_7 \, ,
\\
\xi &\equiv-4 c_4 + c_5 - \frac{3}{2} c_6 - c_7 \, ,
\\
\nu &\equiv c_2 + \frac{5}{12} c_3 + \frac{2}{3} c_4 + \frac{1}{6} c_5 + \frac{1}{8} c_6 + \frac{1}{6} c_7 \, .
\end{align}
Now we will extract the different possibilities for the large momentum dispersion relations. Note that for $N(r) \neq 1$, the general dispersion relation~\eqref{generaldisp} receives the simple modification,
\be 
\omega^2=c_{r}^2 \frac{k_{r}^2}{N^2} +c_{\theta}^2f k_{\theta}^2+f \mathcal{M}^2\, .
\ee
Let us now proceed with that analysis.

\subsection{Solution 1}

To obtain the first dispersion relation we solve $\mathcal{E}_{\theta \phi}$ by setting
\be 
A_0 = - \frac{k_r A_1}{\omega N} \, ,
\ee
and substituting this into $\mathcal{E}_{r \phi}$. That equation then gives one dispersion relation 
\be 
\omega^2 = f k_\theta^2 + \frac{kr^2}{N^2} \, ,
\ee
which is the same as the usual Einstein gravity case. This mode is unaffected by any of the cubic contributions to the action.

\subsection{Solution 2}
Next, as before, we solve  $\mathcal{E}_{\theta \phi}$ by setting
\be 
A_0 = - \frac{k_r A_1}{\omega N} \, .
\ee
However, this time we solve the second factor of $\mathcal{E}_{r \phi}$ to obtain the dispersion relation. The result has the form
\be 
\omega^2 = c_\theta^2 f(r) k_\theta^2 + c_r^2 \frac{k_r^2}{N^2} + f \mathcal{M}^2 
\ee
with
\begin{align}
c_r^2 &= 1 \, ,
\\
c_\theta^2 &= \frac{\alpha r^2 f'' - 3 \beta r f' + 24 \nu (f-1)}{12 \nu r^2 f'' - 3 \beta r f' + 2 \alpha (f-1)}  \approx 1 \, , 
\\
\mathcal{M}^2 &= \frac{2 r^2}{\ell^4 \left((12 \nu r^2 f'' - 3 \beta r f' + 2 \alpha(f-1) \right)} \approx  - \frac{2 r^3}{r_0 \ell^4 (2 \alpha + 3 \beta + 24 \nu)} \, .
\end{align}
Above, for $c_\theta$ and $\mathcal{M}$, to go from the exact expressions to the approximate form, we expand the result for small values of $\ell/r_0$. 

\subsection{Solution 3}
The final possibility for the dispersion relation is obtained by solving the second factor of $\mathcal{E}_{\theta \phi}$. This immediately gives the following
\begin{align}
c_r^2 &= 1 \, , 
\\
c_\theta^2 &= \frac{\alpha r^2 f'' - 3 \beta r f' + 24 \nu (f-1)}{2(\alpha + \xi) r^2 f'' + 2(\alpha + 12 \nu) r f' - (24 \nu + 4 \xi + 6 \beta +6 \alpha)(f-1)} \approx -\frac{1}{2} \, ,
\\
\mathcal{M}^2 &= \frac{r^2}{\ell^4 \left((\alpha + \xi) r^2 f'' + (\alpha + 12 \nu) r f' - (12 \nu + 2 \xi + 3 \beta + 3 \alpha)(f-1) \right)} \approx \frac{r^3}{r_0 \ell^4 \left(2 \alpha + 3 \beta + 24 \nu\right)} \, .
\end{align}
Above, for $c_\theta$ and $\mathcal{M}$, to go from the exact expressions to the approximate form, we expand the result for small values of $\ell/r_0$. 

\subsection{Summary}

The only mode that the EFT captures is that described by solution 1, which agrees with the Einstein gravity case. The additional massive modes are outside the domain of validity of the EFT for the same reasons given in the discussion in the main text. Thus the results for general cubic theories are analogous to the ECG case.  This is not surprising. As we have emphasized in the main text, ECG captures the most general EFT for vacuum gravity with cubic powers of the curvature. Therefore it would have been problematic for the entire EFT program had the general cubic theory provided any essential differences.

\appendix

\bibliographystyle{JHEP-2}
\bibliography{Gravities}
\vspace{1cm}
\noindent \centering
{\color{white} Rocking through the limits, ECG stands strong.\\ Defying claims of instability, proving them wrong.\\ (ChatGPT)}
  
\end{document}
%